# Experimental and modeling methodologies for the analysis of water adsorption in food products. A review

Iván D. Buitrago-Torres 1*, Javier I. Torres-Osorio 2 and Elisabeth Restrepo-Parra 3

1. National University of Colombia, Physics and chemistry department, PCM computational applications; idbuitragot@unal.edu.co
2. Caldas University, Research group on electromagnetic fields, environment and public health; javier.torres@ucaldas.edu.co
3. National University of Colombia, Physics and chemistry department, PCM computational applications; erestrepopa@unal.edu.co
* Correspondence: idbuitragot@unal.edu.co

**Abstract:** The determination of the isothermal adsorption curves represents a mechanism that allows obtaining information on the process of adsorption of water in organic and inorganic materials. In addition, it is a measure to be considered when characterizing the physicochemical and structural properties of the materials. We want to present an approach to the state of knowledge about the methods to characterize seeds and materials associated with food products physically and structurally, and to relate this knowledge to biophysical processes in these materials. This review considers the papers available since 2001 associated with water adsorption studies on seeds and other food products as well as the approach of different authors to to technical and experimental models and processes that are needed for the development of this topic. From these articles the applied experimental methodologies (obtaining samples, environmental conditions and laboratory equipment) and the mathematical models used to give physical, chemical and biological meaning to the results were analyzed and discussed, concluding in the methodologies that have best adapted to the advance of the technology for obtaining isothermal curves in the last years.

## 1. Introduction

Adsorption is a process that is carried out in a system with specific environmental conditions, composed of a solid surface and particles grouped together in a solid or gaseous state. In this process, the particles (adsorbate) are deposited from the solid surface (adsorbent). The characterization of the adsorption process of a certain adsorbate on a specific adsorbent consists of determining the isothermal adsorption curves, which give an idea of the physicochemical mechanisms involved in the process under the established environmental conditions[1]. These curves are obtained from the relationship between the moisture content that a sample has at the time it has reached equilibrium with the pressure of the surrounding environment. This pressure is imposed in the system with values between 0 % and 100 % (figure 1).

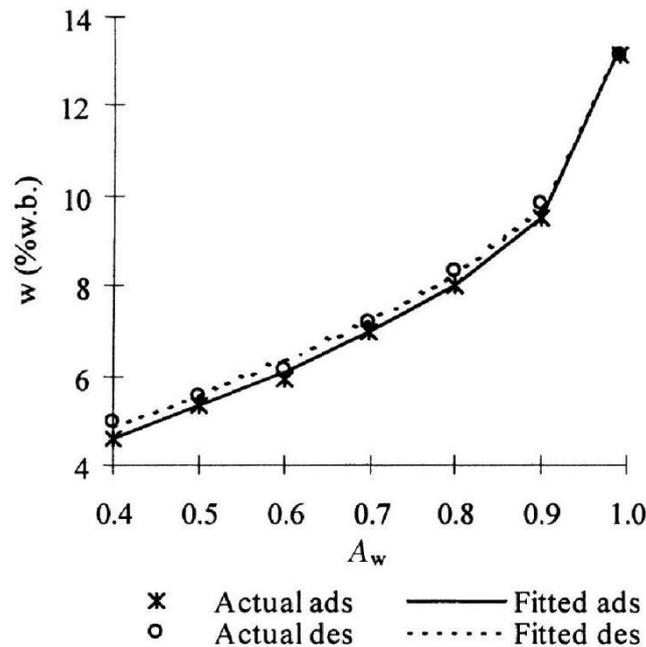

Figure 1. Isothermal curve of water adsorption in bean grains at 25 °C. Obtained from Stencl, & Homola. (2000)[2].

The adsorption processes have been studied since the first mathematical description, attributed to Irwin Langmuir, at the beginning of the 20th century, analyzing the adsorption of different gases on glass, mica and platinum surfaces [3]. This process was described as a chemical reaction that allowed the generation of an adsorbed product from an adsorbate and an adsorbent. One of the first refinements made to the Langmuir's theory involved the mathematical description of adsorption as a process of deposition of gaseous particles in the form of consecutive layers on the solid surface and was developed by Brunauer, Emmet and Teller (BET theory) [4].

The current proliferation of different models for the description of adsorption processes has also allowed a wide variety of applications of these phenomena, such as the removal of ions from contaminating substances in aqueous solutions [5]–[7], purely theoretical analysis describing these processes in different products from a thermodynamic point of view [8]–[11] and in the conservation and shelf life of food products [12]–[16].

Becasue of the use of adsorption isotherms in the study of food products, the current document proposes a review of high-impact articles, in order to identify an experimental methodology and a modeling methodology, applicable for future studies.

## 2. Materials and Methods

The bibliographic search was carried out in two stages, using the secondary source "Web of Science" (WoS) during September 2020. For the first part, the terms "water", "moisture", "adsorption", "isotherm" and "seeds" were established as descriptors in the TOPIC search section. The search equation was refined to exclude results concerning the adsorption of heavy metals and dyes because these studies are alien to the topic of this review. The search yielded 223 results from which, the 30 most referenced were taken being cited between 12 and 278 times and published between 2001 and 2017.

The second part of search, in the same source and with the same descriptors, was biased between the years 2017 and 2020, and all the articles obtained, 18 in total, were retrieved to give a perspective of the reports generated in recent years. The other articles referenced in this review correspond to complementary information concerning theoretical and experimental aspects, models and processes that are required for the proper development of the document.

## 3. Results

The review of the methodologies for obtaining and analyzing the isothermal adsorption curves presented was carried out from a division of the process into two segments. One is the analysis of the experimental methodology and the other is the inspection of the modeling and the data analysis. The experimental section addresses the arrangement of the experimental setup (temperature, pressure and moisture content determination) and the treatment that the samples have before the experiment. In the data modeling and analysis section, the main characteristics of the analysis models most used in the literature and the characteristics of some models are considered.

### 3.1. Experimental Methodology

The experiment to obtain isothermal adsorption curves consists of exposing the sample (the adsorbent) to a controlled environment of pressure and temperature until isobaric and reactive equilibrium conditions are reached (constant mass); then, it is necessary to analyze both, the arrangement of the experimental system as the way to achieve the pressure and temperature conditions demanded according to the study. It is important to consider the nature of the samples used as adsorbents, since in many cases they are subjected to strict manipulations before the experiment.

3.1.1. Arrangement of the experimental system

The standardized system for obtaining adsorption isotherms in foodstuffs was established through two complementary projects, COST Project-90 and Project 90 bis (COST: Cooperation in the field of scientific and technical research in Europe) carried out consecutively in 1983 and 1985, summarized and published by Wolf and Speiss [17]. The experiment consists of positioning the sample in an isolated system that allows controlled temperature and vapor pressure conditions to be. Methods for obtaining pressures and temperatures will be discussed later. Figure 2, adapted from [17], shows the graphic description of the experiment established as a standard for obtaining isotherms in food products. In the articles reviewed, the experimental methodology is like the one proposed in the standard, making small changes in components that allow it to keep the same principles of preserving environmental conditions and subjecting the sample to these conditions.

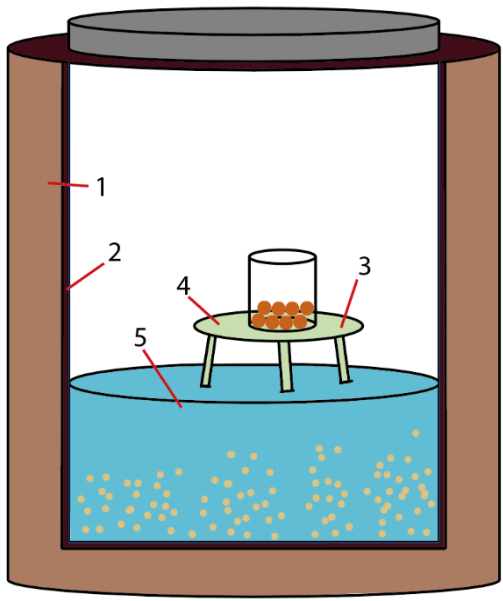

Figure 2. Detail of the standardized experiment in COST proect-90. 1: Thermal water bath. 2: Closed container. 3: Weigh bottle with mobile stopper. 4: Tripod. 5: Saturated salt solution. Adapted from [18]. Source: The authors.

To develop the above, it is necessary to name the methodology presented in the vast majority of studies [19]–[28] and consists of the use of small uncovered containers that contain the samples. These containers are arranged in desiccators that supply the need for a tripod, with the saturated solution at the bottom of them as outlined in figure 3. The desiccators are hermetically sealed and the whole system is immersed in an incubator that allows keeping the temperature controlled.

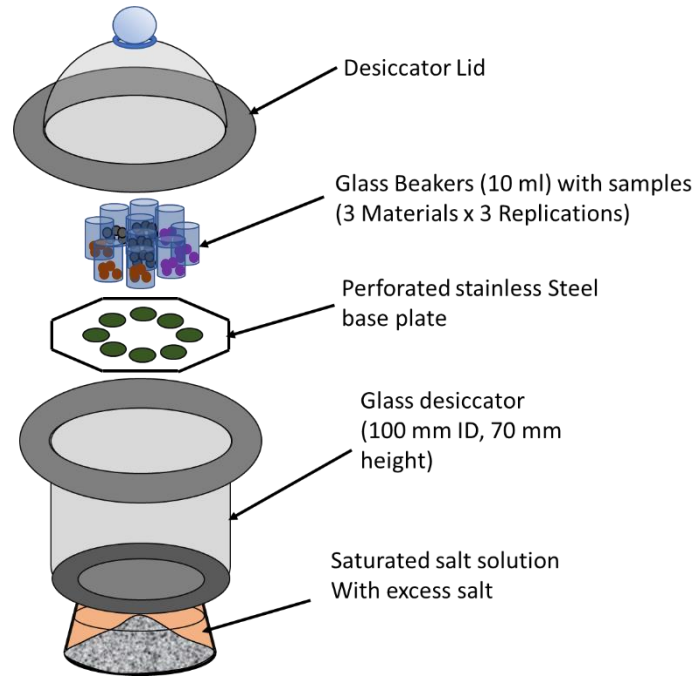

Figure 3. Adsorption experiment diagram using a desiccator. From top to bottom: Top lid, containers containing samples, perforated base, desiccator, saturated salt solution. Adapted from [29].

There are also a few lesser-used methods that allow the recreation of the same original dynamics as the standard. In their studies Torres[30] and Vashisth[31] used a system that replaces the use of desiccators with a compact alternative by unifying containers, saturated solution and the tripod in a single closed Petri dish that contains the samples in porous bags that allow the flow of water vapor and a mesh that prevents contact of the seeds with the solution (figure 4).

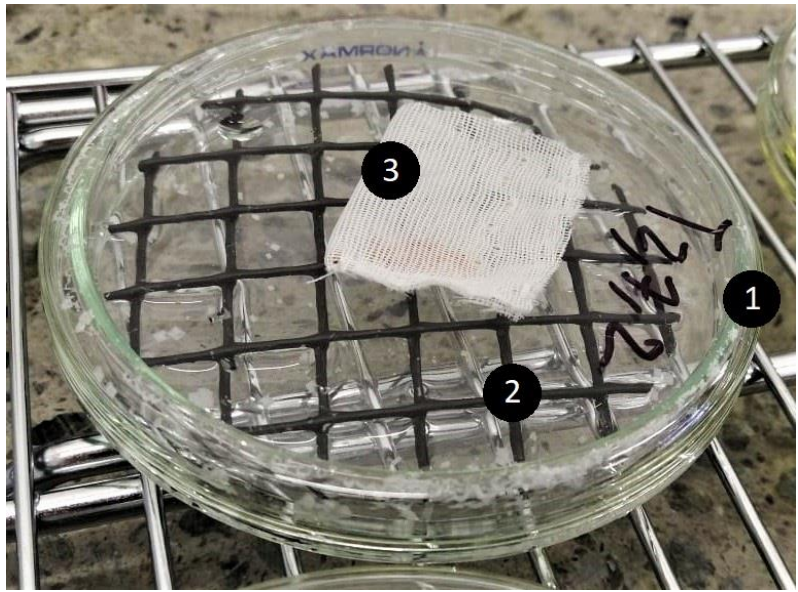

Figure 4. Top view of the simplified experimental setup. 1: Petri dish to be placed inside an incubator, with saturated salt at the bottom. 2: rigid plastic base with perforations that separates the samples from the solution. 3. porous mesh bags containing the samples from the experiment. Source: The authors.

It is highlighted the fact that the methodologies used in the literature are based on the adaptation to the original standardization and although there are differences in the execution, the same principles and experimental objectives remain.

- Temperature management and selection

The selection of the temperature of the experiment is established according to the objective of the investigation. The tendency found in the selection of temperatures is due to the fact that the investigations are mainly focused on the analysis of the behavior of the seeds in the face of storage conditions, and the search of information that allows the prevention of damage to the product due to alterations in its long-term nutrients[32],[33] and the effect of water activity by dehydration [34].

Experimentally, the temperature values selected for the study of the samples are achieved using thermally isolated systems, thermal baths, ovens or incubators with or without air circulation. Figure 5 presents a relationship of the number of times certain temperature values were used in the reviewed studies. In these, ovens or incubators are used since it is part of the standard[17], except for only two particular cases presented below.

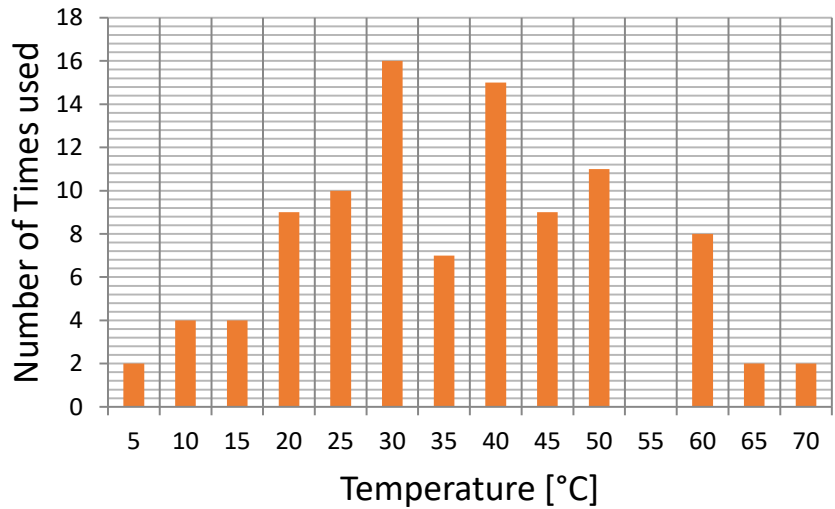

Figure 5. Frequencies of using temperatures in experiments. Source: The authors.

The justification presented by Togrul and Arslan[35] not to use a regular incubator lies in the very nature of a system created by themselves to obtain isothermal adsorption curves by using a chamber made of Cr-Ni. The publication they present does not specify the way to reach the temperature, but they highlight their ability to maintain it. In Zeymer's case [36], it is explained that the process of establishing the temperature is achieved through the use of a biochemical oxygen demand chamber (BOD chamber) as in the study of Isquierdo[37].

- Pressure conditioning

The different pressures in each of the closed systems can be modulated with the use of solutions saturated with certain salts, which in a standardized way provide a vapor pressure within the system. The pressure in the system is strictly related to the amount of gas phase adsorbate to which the sample is exposed and is, together with the temperature, the independent input variable of the experiment. Saturated saline solutions are mixtures of distilled water and chemical compounds (salts) with low amounts of impurities. The salts must be in a sufficient quantity for the mixture to reach or even exceed its saturation point[38]. It has been shown that the use of saturated solutions of different salts in a closed system allows specific pressure conditions to be reached in each system by varying the salt in the solution.

It is possible to describe the pressure of the system by means of two physical quantities, the relative humidity (given in percentage) that indicates the proportion of water molecules in the environment and has an operating range from 0% to 100%. Interval that corresponds to the variation between a totally dry environment and a water-saturated environment, respectively. On the other hand, the water activity, which is the corresponding dimensionless decimal of relative humidity and has values between 0 and 1, both magnitudes are related by

$$a_W[dimensionless] = R\text{elative humidity}[\%] / 100\%, \qquad (1)$$

It was possible to determine that in the articles reviewed in this document, relative humidities were generated from 5% to 98%; however, it is necessary to use the activity of water as a pressure measure ($a_W$) to carry out the corresponding analyzes.

For the selection of salts the reports unanimously refer to the tables provided by Greenspan [18] in 1977, that show the relationship between the vapor pressure of the system and the saturated solutions depending on the temperature. Greenspan summarizes the possible use of a total of 28 salts that can be used to prepare saturated solutions with distilled water and that allow a range of relative humidity between $(1.63 \pm 0.64)$ % and $(98.77 \pm 1.1)$ %. Table 1 shows values corresponding to the first and last column elements in the Greenspan tables, to give a notion of the relative humidity range that can be achieved in an isolated system only with the use of saturated salts at different temperatures.

**Table 1**. Relative humidity of Cesium Fluoride and Potassium Chromate for different temperature values [18].

|        | Relative humidity [%] | |
|--------|-----------------|-------------------|
| T [°C] | Cesium Fluoride | Potassium Chromate |
| 5      | 5.52            | *                 |
| 10     | 4.89            | *                 |
| 15     | 4.33            | *                 |
| 20     | 3.83            | *                 |
| 25     | 3.39            | 97.88             |
| 30     | 3.01            | 97.08             |
| 35     | 2.69            | 96.42             |
| 40     | 2.44            | 95.89             |
| 45     | 2.24            | 95.5              |
| 50     | 2.11            | 95.25             |
| 55     | 2.04            | *                 |
| 60     | 2.03            | *                 |
| 65     | 2.08            | *                 |
| 70     | 2.2             | *                 |
| 75     | 2.37            | *                 |
| 80     | 2.61            | *                 |

* Reports for these temperature values are not available in the literature.

In some publications, instead of saturated saline solutions, other alternatives are used to achieve relative humidities in the system in a controlled way. In their study, Togrul et al. [35] present in the system created by them to obtain the isothermal curves, a variation of the pressure based on the modulation of the amount of silica gel instead of saturated solutions. As another method for pressure control, two articles present methods of humidification by water spray [39] and silica gel spray [40] regularly measuring system pressure to keep it constant. As a last alternative method to the use of saturated saline solutions, Al-Muhtaseb [41] cites the use of $H_2SO_4$ solutions in different concentrations that allow, through the dissociation of their ions, to expel controlled amounts of water to the system.

Due to the complexity of controlling the relative humidity in a system, even if it is isolated, the use of saturated salt solutions, currently remains the most widely used technique due to its easy use and its tradition in experiments of this nature.

- Determination of moisture content

In addition to the COST Project-90 standard that indicates how to proceed experimentally when carrying out a water adsorption study on different samples, there is another standard frequently cited by the articles referenced here and it consists of the methodology to perform the measurement of the dry mass of the samples when these are coming from fruits. The process of determining the moisture content in fruit-derived products is standardized by AOAC (Official Methods of Analysis) and the 1980 proposal is presented here [42]:

"22.013 Moisture in nuts: Spread (5-10) g of the prepared sample as evenly as possible on the bottom of a metal plate approximately 8.5 cm in diameter with a tight-fitting lid, measure mass and dry for six hours at (70 ± 1) ° C under pressures less than 100 mmHg (13.3 kPa). The metal plate must be in direct contact with the metal shelf of the oven. During drying, allow a stream of dry air in the oven (approx. 2 bubbles / s), passing it through $H_2SO_4$. Replace the lid, cool the plate in a desiccator and measure the mass. Ignore any change in oven temperature during the first stage due to evaporation of water."

Contrary to the AOAC standard, the reviewed articles used ovens with temperatures between 103 ° C and 110 ° C in time intervals between 3 h and 28 h, depending on the mass of the sample. Two articles have relatively low temperatures for determining the initial moisture content and dry mass. Kaya [43] and Moreira [44] used an experimental temperature of 70 ° C for six hours for sesame seeds and 36 hours for layers of loquat and quince respectively.

Despite the difference in temperatures and exposure times used in different studies, the objective in all cases is to allow a total evaporation of moisture in the sample, avoiding important morphological changes due to exposure to high temperatures by extended time. This change in mass due to the loss of moisture makes it possible to measure the final moisture content in the samples once an equilibrium is reached in the value of the mass, a situation that occurs when a change of less than 0.001 g in a time range of 24 h according to the standard of the gravimetric method [42].

Two methods are used to express the moisture content of the samples: the wet basis and the dry basis wet content [45]. The moisture content on a wet basis ($M_W$) is the percentage rate between the mass of water and the total mass of the product

$$M_W = \frac{W_W}{W_t} \times 100 = \frac{W_W}{W_W + W_d}, \qquad (2)$$

where $W_d$ is the mass of the wet material, $W_t$ is the mass of the whole sample and $W_W$ is the water mass. The dry base is a percentage expression of the ratio between the mass of the water content and the mass of the dry material

$$M_d = \frac{W_W}{W_d} \times 100, \qquad (3)$$

This dry basis is used in adsorption studies because it describes the mass of the sample in relation to moisture content. Thus, it is possible to measure the moisture content on a dry basis in the moisture balance and after drying the sample, to determine the total amount of moisture adsorbed.

3.1.2. Sample obtaining

As described in the methodology section, the search equation was proposed with the objective of obtaining publications concerning seed adsorption studies; however, the results showed that not only intact seeds are the object of study, but in many cases, they undergo some treatment or procedure before their study. For this reason, this section is divided into studies that used samples obtained from some part of the fruit and those that used the seed.

- Samples from the fruit

It is possible to identify in the literature studies in which, the fruit of some species is used in its entirety. Vankatachalan [46] looked for the isothermal characteristics in walnuts without any treatment like Goneli [27] in the study on pearl millet and Rodrigues [47] *et al.* in peppers. There are also studies that use the fruit almost entirely with minimal changes in its morphology as proposed by Lahsasni, **Kouhila and Mahrouz** [48] in the study in 2004, on prickly pear fruit. In this work, the fruit was cut until achieving a mass of 1 g. Then in 2015 it was presented information on the adsorption in nopal fruit [21], for its study the fruits were washed in water, peeled, and the pulp was manually separated from the seeds, the latter were stored until use.

Another method of obtaining the sample in which parts foreign to the seed are incorporated was that carried out by Arslan and Togrul [40] who used crushed chili peppers without further treatment. Moreira [44] studied the hygroscopic behavior of loquat and quince, separating the quince from its shell and cutting both samples into 3 mm layers. Studies based on the use of derivatives of fruit pulp, such as the pumpkin parenchyma carried out by Mayor [49] and of Chawla with tomato [20], identify processes to obtain a final product from the pulp of the fruit. As a food product and without being properly identified as fruits, potatoes are an investigative objective [50]; then, McMinn [19] analyzed in 2003 the hygroscopic behavior of this without any type of treatment. Al-Muhtaseb presented a thermodynamic analysis as a continuation of a previous study [22] in which, the isothermal adsorption curves for potato starch were determined, for this study, starch powder was extracted, which served as a sample in the adsorption and desorption experiments.

- Samples from the seed

Since in different instances it is pertinent to study the storage characteristics of seeds in their natural state [51] or after some type of priming to know the modification of the sample before the treatments carried out, this section includes most of the articles reviewed for this document.

- Whole seeds

The use of whole seeds has been reported in multiple products such as melon, which was studied by Aviara [24]. Among other sedes analyzed, without any treatment, were peppers [52], hemp[53], black chia[54], avenda nuda[55] and urunday seeds[37]. There are a series of studies carried out at the University of Rio Grande do Sul on the pinion (A. angustifolia), among which, the study by Cladera-Olivera stands out [56] in which the use of raw seeds of Pinion commercial type is described.

In 2006, Kaya presented a research [43] on sorption in sesame seeds. For this study, the seeds were used in three stages of the same treatment, these were sieved and immersed in water at 18 °C for 12 h, then dried (first group of samples) and passed through a mechanical peeler to remove the shell from the seed, the other impurities were removed from the seed by immersion in salt and then, washed with plenty of water, centrifuged (second group of samples) and then roasted on a rotisserie at 150 °C for 100 min. These last seeds contemplate the third sample studied.

- Kernel

Concluding with the second part of the study previously exposed by Mayor [49], In addition to the parenchyma obtained from the pumpkin, the study was also carried out on samples obtained from the seeds, which were cut into slices, without specifying their average dimensions, but with a total mass of 2 g. In some studies, seed treatment is given simply by selecting one of its macroscopic divisions as the study sample, among these is the study by Ali Abas Wani [57], who presented an study of two types of watermelon. Similarly, Australian orchid seeds were studied [58] extracting the kernel, washing them manually, drying them in ambient conditions for a week and finally storing them in polyethylene until use. A study on walnut [35] proposes the extraction of the internal seed, after the crushing of the testa and the use of the internal one to create cubes with 1 cm on each side. This process is repeated in the literature in a more recent study by Sahu [59] in 2018.

- Powders and others

Even though the fruits and seeds have the main interest in the study of the appropriate conditions for their storage, powders derived from food products have been equally recurrent in the literature because the spraying process in some cases allows better regulation, in addition to the fact that these are also a food product.

Returning to the studies carried out at the Rio Grande do Sul University, Porto Alegre, Brazil, on adsorption in pinion, three relevant papers are identified, one led by Cladera-Olivera [60], and another by Thys [61]; in both reports, they focus on the process used to extract powder. The first exposes the process from the remnant after removing the two outer layers of the seeds, while the second exposes a similar process using the entire seed. In the third report by the same research team [62] the same methodology was used to obtain the powder previously used [60], adding a few extra steps to obtain pinion microcapsules from the powder.

Another study that is based on the sampling of components foreign to specific parts of the fruit or seed uses tapioca as a precursor of the sample [26] This is obtained from cassava flour, coconut milk, egg yolk and sucrose. The obtaining process is described step by step in the report to finally present the isotherms of the sample. Finally, the most recent studies carried out by Hoyos-Leyva [39] present the preparation of an aggregate from taro corn from which starch was initially obtained and then an aggregate after processing by dry spray.

It is then possible to identify different approaches to obtaining the sample from a food product for water adsorption studies. The main goal of these studies is to predict the behavior of the sample under storage conditions and to try to optimize these conditions to avoid the degradation of the product by unwanted reactions at the time of its conservation. However, depending on the specific objectives, working conditions and application interests of the research, different ways of handling the samples can be distinguished before their disposal in the experimental setup.

*3.2. Methodology of data modeling and analysis*

Adsorption studies are mathematically based on modeling using equations that have been proposed, refined, and used since the first hypothesis developed by Irwin Langmuir in 1918 [3] to this day. These models seek to give a description of the physisorption process (adsorption that does not involve chemical reactions) [63] from non-linear adjustments of the curves obtained for the quantity of water measured in a sample with respect to the relative humidity of the system, for a given temperature. The measurement of the humidity in the samples must be recorded once an equilibrium is reached in their mass, a situation that occurs when a change of less than 0.001 g is registered in two consecutive measures within 24 h, if the resolution allows it, according to the standard of the gravimetric method [42].

In several cases, the models allow some of its variables to be related to physical properties of the system, such as surface area [4],[64], the volume of moisture in the sample [65],[66] and the volume adsorbed by the monolayer [67],[68]. Currently, it is possible to distinguish two groups of adsorption models: the independent and dependent on the temperature of the system. Some temperature independent models are Freundlich [69], BET [4],[68], Harkins-Jura [70], Oswin [65], GAB [67],[68],[71], D'Arcy-Watt [64], Smith [72], Halsey [66], Henderson [73], Chung-Pfost [74] and Chirife [75]. The second group corresponds to models based on the independent ones with the addition of a temperature term, some of the temperature dependent models are modified Chung-Pfost, modified Hasley, modified Oswin, and modified Henderson [76].

Although the number of models historically reported according to Goneli [27] is approximately 200, Figure 5 shows the 29 models that were used in the reviewed articles (generally more than one per article) and the number of times each was best adjusted to the study in question. The six models that were most applied (GAB, BET, Oswin, Henderson, Halsey and Peleg) will be briefly explained to contemplate their main characteristics and the information on the physical-chemical process that they provide according to their parameters.

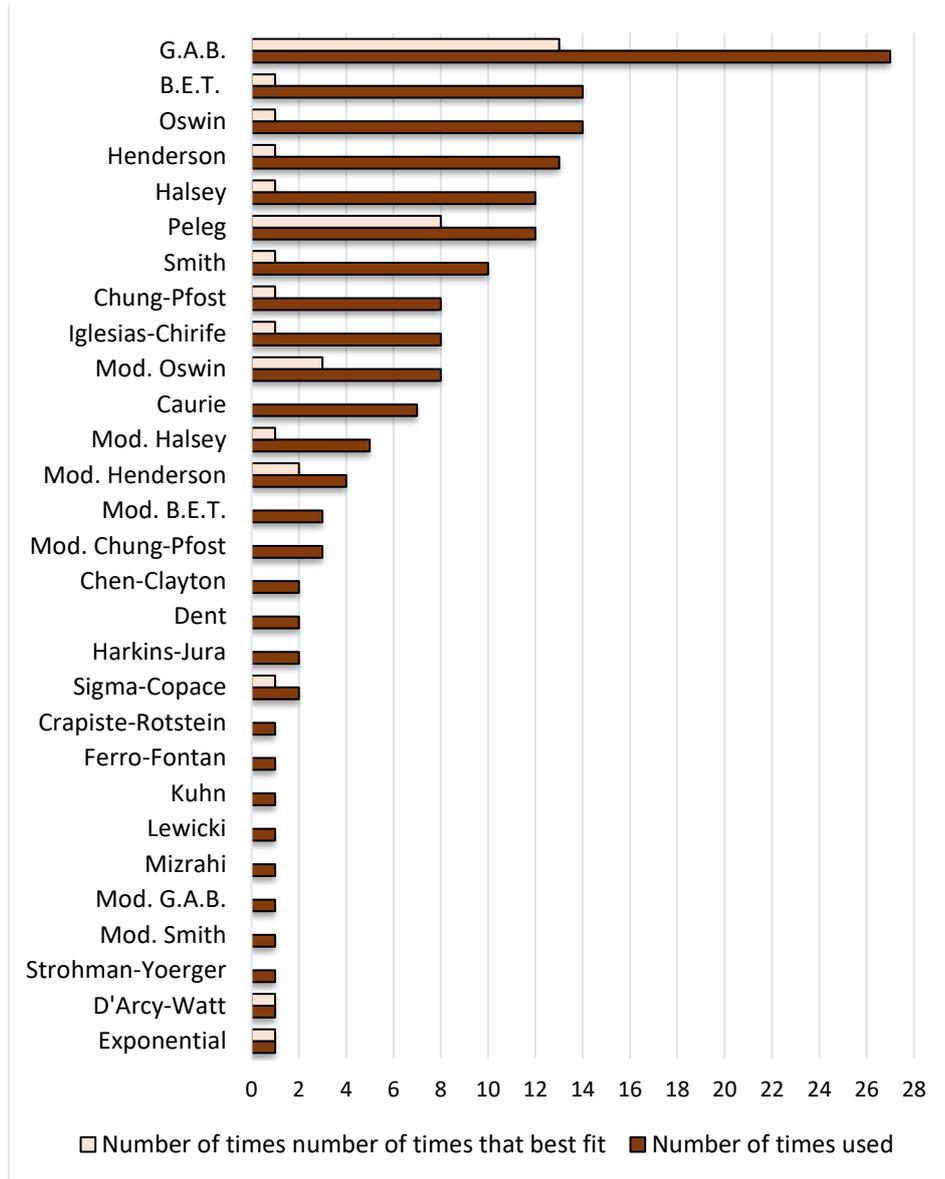

Figure 6. Adsorption models, their frequencies of use and times that were the best fit.

3.2.1. GAB model

The predominant model in the literature for the description of the behavior of equilibrium moisture content in food products is the GAB model. This model is proposed separately by Guggenheim, Anderson and de Boer as a refinement of the BET model. It considers the interaction energy of the multilayers identical to the energy of the second layer of the adsorbent in the adsorbate but different from its liquid phase. This refers to an infinite assumption of multilayers in the model. The GAB model is expressed according to Berg [77] as

$$M_W = \frac{M_0 C K a_W}{(1-K a_W)(1-K a_W + C K a_W)}, \qquad (4)$$

where $M_W$ corresponds to the equilibrium moisture content expressed in kg over 100 kg of dry solid, $M_0$ corresponds to the moisture content in the monolayer, C (Guggenheim constant) and K are the adsorption constants that are expressed as

$$C = c_0 \exp((H_0 - H_n)/RT), \qquad (5)$$

$$K = k_0 \exp((H_n - H_i)/RT), \qquad (6)$$

where $c_0$ and $k_0$ correspond to entropic accommodation factors and the values of $H_0$, $H_n$ and $H_i$ are the molar sorption enthalpies of the monolayer, the last multilayer and the liquid bulk, respectively.

In the studies analyzed for this review, the GAB model was the best fit for potatoes[19], dehydrated prickly pear[21],[48], pumpkin seeds[49], amaranth grains[78], nut[46], tarragon petals[79], loquat, quince [44], chia mucilage[80], taro starch [39], oak wood[81], tamarind mucilage[82] and chironji kernel[59].

### 3.2.2. BET model

The BET model, named after the proponents (Brunauer-Emmet-Teller), provides an effective method for determining the amount of water bound at specific polar sites of the adsorbent [83],[84]. The theoretical development in the BET model involves 3 important assumptions: The condensation ratio of the first layer is equal to the evaporation ratio of the other layers, the bond energy of all sites on the surface is the same and finally the bonding energy of the multilayers is equal to the bonding energy of the monolayer. Despite these unrealistic assumptions about constant energy interaction in all the adsorbate layers involved, the model continues to be a good starting point for proposing new models [68],[84],[85]. BET model is described by

$$M_W = \frac{M_0 C a_W}{(1-a_W)(1+(C-1)a_W)}, \qquad (7)$$

where $M_0$ it is the moisture content in the monolayer, which indicates the maximum amount of adsorption allowed in the monolayer before the multilayer adsorption begins, C is an energy constant related to the difference in net adsorption heat between the monolayer molecules in the other layers.

The BET model provided the first notion of a multilayer model at the time of its development; nevertheless, its difficult adjustment to a full range of water activities (since it is only linearly adjustable to low activities $a_W < 0.5$) makes it useful mainly in determining surface area [66]. Among the articles reviewed, the BET model presented the best fit with the experimental data of water adsorption in nimbus [86].

### 3.2.3. Oswin model

It is an empirical model developed from the series expansion of the isothermal curves typically obtained in adsorption processes, developed in 1946 by Oswin [65]. Given that the model was developed from a mathematical proposition based on the experimental data of adsorption on various products (apple, dehydrated carrot, dry eggs, paper, cigarettes, potatoes) applied to a generic sigmoid equation, the model provides an appropriate adjustment to the isotherm of certain products but does not allow characterizing any physical or chemical properties of the adsorbent. The Oswin model is described by

$$M_W = C \left(\frac{a_W}{1-a_W}\right)^n, \qquad (8)$$

where $C$ and $n$ are constants for being determined.

Oswin model is presented as the best fit for watermelon seeds [57]. Moreover, the modifications of this model are also the best adjusted for lettuce seeds [36], chili pepper "malagueta" seeds [52] and wheat [87].

### 3.2.4. Henderson model

Henderson equation is expressed according to [73] as

$$M_W = \left(-\frac{\ln(1-a_W)}{C}\right)^{\frac{1}{n}}, \qquad (9)$$

where $C$ and $n$ are constants for being determining.

The Henderson equation, in addition to the reviewed products for which it proved to be a good fit, has been shown to be the isotherm that describes adsorption on red clover petals [2], farina and semolina [28] and its modified version presents the best fit for tomato pulp data [20] and pearl millet grains [27].

### 3.2.5. Halsey model

The Halsey model is a model that proposes the description of adsorbent multilayers on the surface of the adsorbate assuming that the energy decreases in inverse proportion to the distance. This equation has generally been found to describe the behavior of products containing starch [88], [89] and it is expressed as

$$M_W = M_0 \left(-\frac{A}{RT\ln(a_W)}\right)^{\frac{1}{n}}, \quad (10)$$

where $A$ and $n$ are constants, $R$ is the universal gas constant, $T$ is the absolute temperature of the system and $M_0$ is the initial moisture content. This equation however was modified by Iglesias and Chirife [90] to become

$$M_W = \left(-\frac{C}{\ln(a_W)}\right)^{\frac{1}{n}}, \quad (11)$$

where $C$ and $n$ are constants.

This model allowed to satisfactorily describe the behavior of water adsorption in sesame seeds [43] and the modified model, that of melon and cassava [24].

Even though there are dozens of models of adsorption isotherms in the literature that have made it possible to satisfactorily describe the adsorption of both water and other adsorbates in different types of adsorbents, it is important to note that historically certain models have been a common reference for the study of some specific products, and the GAB, BET, Oswin and Henderson models are a good starting point for the isothermal characterization of any food product.

**4. Discussion**

The detailed review of the most cited articles in the literature, according to our search equation, allows to focus on the different stages and characteristics that involve the sorption experiment in food products. In this process of analysis of the studies, it was possible to discern several critical moments that involve methodological choices for the correct execution of the experiments. These moments are the selection of the experimental setup, the control of pressure and temperature, the selection of the sample and the selection and analysis of the mathematical model with which the experimental data is described.

The experimental set-up, despite evident differences in its implementation, maintains on all occasions what is established by the COST Projet-90 standard. This standard proposes the use of an insulated container at a controlled temperature in which the sample is maintained and a constant humidity is established. Although saturated salt solutions and furnace-immersed desiccators are used in most studies to achieve this proposal, new methodologies have been implemented in order to facilitate repeatability and increase the efficiency of experiments following the same experimental principles.

Even though the temperature must be established and constant for each isothermal curve obtained independent of the sample under study, the effect on the change of said temperature in the shape of the curve obtained is not considered on all occasions. Whether one or more temperatures, in all cases these are kept around the same values in which samples are stored under normal conditions. The most widely used temperature values in the literature are 30 and 40 ° C, but studies are found from 5 to 70 ° C, avoiding the freezing points of water and cell death due to high temperatures. In addition to the values used in the reviewed studies, it was also possible to obtain the experimental methods used to achieve these temperatures. Furnaces are used in most studies, being the most common and simplest method. However, it is possible to find some variations such as thermal water baths and the use of BOD chambers.

Pressure is the second environmental parameter involved in the experiment and is expressed in terms of water activity or relative humidity. The values vary between the absence of water in the system (0% relative humidity) and total flooding (100% relative humidity). The way to achieve these pressures within the isolated chamber (usually dissectors) is by saturated saline solutions. These solutions and the pressure values to which the system is subjected were filed by Greenspan [18] and in all the cases in which they are used, said study is cited.

Regarding the samples to be studied, despite having the same common origin (food products), the papers reviewed here point to many ways in which it is possible to treat and manipulate them before beginning the experimental phase. This is how we found samples based on the seeds without any alteration, seed shells, kernel, part of the fruit, part of the shell, flour based on the initial fruit, cylinders and cubes from the fruit and others, which give a notion of the number of ways in which it is possible to store these depending on the use that will be given.

After the experimental phase, with the respective data recorded, all the studies proceed to find the best fit of these data among various mathematical models available in the literature. Each of these models presents different parameters that, according to their nature, serve as references to characterize physical properties of the system and dynamic properties of the reaction that is being carried out. Within these models, the most widely used (GAB, BET, Oswin, Henderson, Halsey and Peleg) were explained to give a notion of the properties that allow describing and it was determined which properties were the best fit and in which products.

Finally, it is possible to highlight the particularity that each of the studies reviewed here present both experimentally and methodologically due to situations that involve both instrumental capacity and the objectives of each study. Although these studies always seek to stay within the margins established by the experimental standard, there are so many variations of the original proposal that it is necessary to insist on remaining faithful to the regulations of the standard.

**5. Conclusions**

Although there are studies that involve the modification and improvement of the experimental process required to determine the isothermal adsorption curves in food products, no comparative studies are identified that expose evaluations on the way to arrange the samples in their respective containers, the geometry and arrangement of the samples themselves, the chemical composition and its possible influence of the materials used in the process and the intrinsic characteristics in the surface morphology of the samples.

Regarding the modeling of the experimental results, it is determined that, despite the existence of hundreds of models in the literature, few have more accurately described the isothermal adsorption curves in food products. The analysis of the articles referenced here yielded a high success rate as the best model to the GAB model, however, the evaluation of multiple models in the same experiment must be considered for comparative and statistical reasons.

The experimental process to obtain isothermal adsorption curves continues to be done in the framework of the established standards, however, this same tradition has closed the way for it to be possible to find studies from other fields such as computational simulation, which would allow the development of the experiments at a lower cost in run time, supplies and labor.

Finally, given the wide range of types of samples to be used as adsorbent in adsorption studies in food products, it is possible to propose the study of the hygroscopic behavior of other parts of the plants of interest such as leaves, roots or stems.

**Funding:** This research received no external funding.

**Institutional Review Board Statement:** Not applicable.

**Conflicts of Interest:** The authors declare no conflict of interest.